\documentclass[Physsubmission, Phys]{SciPost}

\hypersetup{
    colorlinks,
    linkcolor={red!50!black},
    citecolor={blue!50!black},
    urlcolor={blue!80!black}
}

\usepackage[bitstream-charter]{mathdesign}

\newcommand{\bp}{\mbox{\boldmath $p$}}
\newcommand{\bq}{\mbox{\boldmath $q$}}

\DeclareSymbolFont{usualmathcal}{OMS}{cmsy}{m}{n}
\DeclareSymbolFontAlphabet{\mathcal}{usualmathcal}

\begin{document}

\begin{center}{\Large \textbf{
Central dileption production in proton-proton collisions
with rapidity gap and with forward protons\\
}}\end{center}

\begin{center}
A. Szczurek \textsuperscript{1,2},
B. Linek\textsuperscript{2} and
M. Luszczak\textsuperscript{2$\star$}
\end{center}

\begin{center}
{\bf 1} Institute of Nuclear
Physics, Polish Academy of Sciences, ul. Radzikowskiego 152, 
PL-31-342 Krak{\'o}w, Poland
\\
{\bf 2} College of Natural Sciences, Institute of Physics,
University of Rzesz\'ow, ul. Pigonia 1, PL-35-959 Rzesz\'ow, Poland
\\
* antoni.szczurek@ifj.edu.pl
\end{center}

\begin{center}
\today
\end{center}


\definecolor{palegray}{gray}{0.95}
\begin{center}
\colorbox{palegray}{
  \begin{tabular}{rr}
  \begin{minipage}{0.1\textwidth}
    \includegraphics[width=22mm]{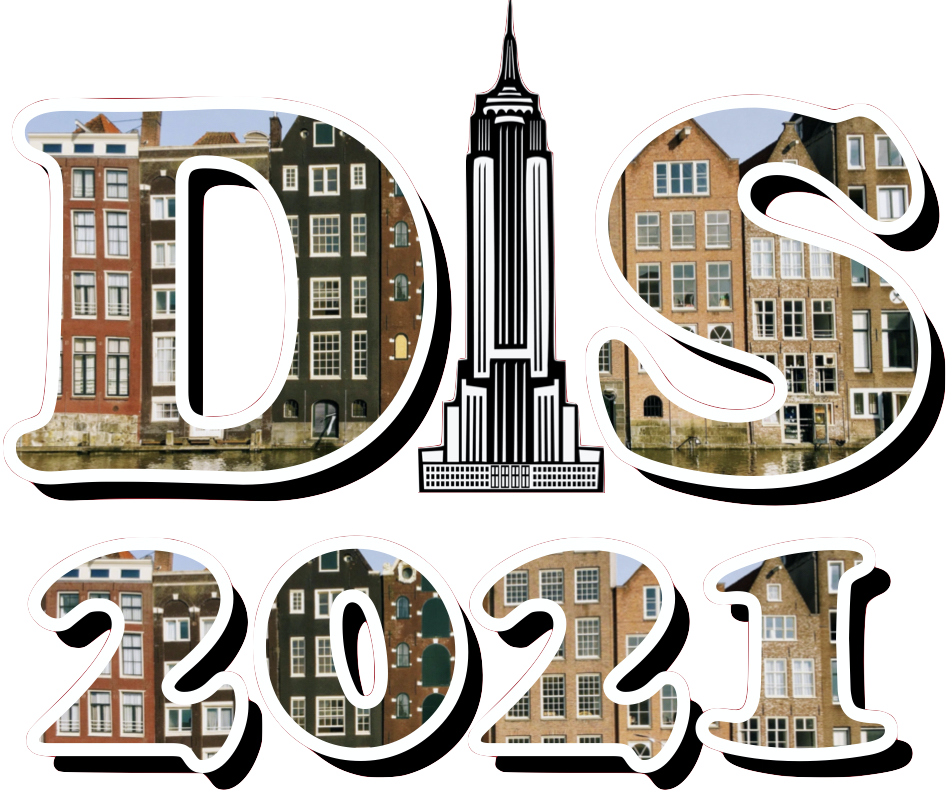}
  \end{minipage}
  &
  \begin{minipage}{0.75\textwidth}
    \begin{center}
    {\it Proceedings for the XXVIII International Workshop\\ on Deep-Inelastic Scattering and
Related Subjects,}\\
    {\it Stony Brook University, New York, USA, 12-16 April 2021} \\
    \doi{10.21468/SciPostPhysProc.?}\\
    \end{center}
  \end{minipage}
\end{tabular}
}
\end{center}

\section*{Abstract}
{\bf
We discuss photon-photon fusion mechanisms of dilepton production 
in proton-proton collisions with rapidity gap in the main detector 
and one forward proton in the forward proton detectors. 
This is relevant for the LHC measurements by ATLAS+AFP and CMS+PPS. 
Transverse momenta of the intermediate photons are taken into account
and photon fluxes are expressed in terms of
proton electromagnetic form factors and structure functions.
Both double-elastic and single-dissociative processes are included
in the analysis. Different parametrizations of the structure functions
are used.
Some differential distributions are presented.
Some differences with respect to the results without proton
measurement are discussed.
}

\vspace{10pt}
\vspace{10pt}

\section{Introduction}
\label{sec:intro}

Only recently the CMS collaboration \cite{CMS} and very recently 
the ATLAS collaboration \cite{ATLAS} presented results with at least 
one proton measured in forward direction. The experimental aparatus 
allows to measure only very forward protons.
In theoretical calculations one has to impose experimental limits on 
so-called $\xi$-variables (longitudinal momentum fraction loss) 
\cite{CMS,ATLAS}.

The results presented here (DIS2021) is based on our recent preprint
\cite{SLL2021}.
In our calculations we use the formalism developed in
\cite{SFPSS2015,LSS2016,LSS2018}, which allows to calculate 
the cross section differential also in $M_X$ or $M_Y$, 
masses of the excited proton remnants.
In \cite{FLSS2019,LFSS2019} it was discussed
how to calculate gap survival
factor which is related to emission of (mini)jets produced in a DIS
process associated with $W^+ W^-$ and $t \bar t$ production, respectively. 
We shall repeat such a calculation also here for $\mu^+ \mu^-$ production.
The absorption for double-elastic contribution was studied 
e.g. in \cite{LS2015,LS2018} using the momentum space formalism.
The impact parameter approach can be found e.g. in \cite{DS2015}.

\section{Basic formalism}

\begin{figure}
\begin{center}
\includegraphics[width=4cm]{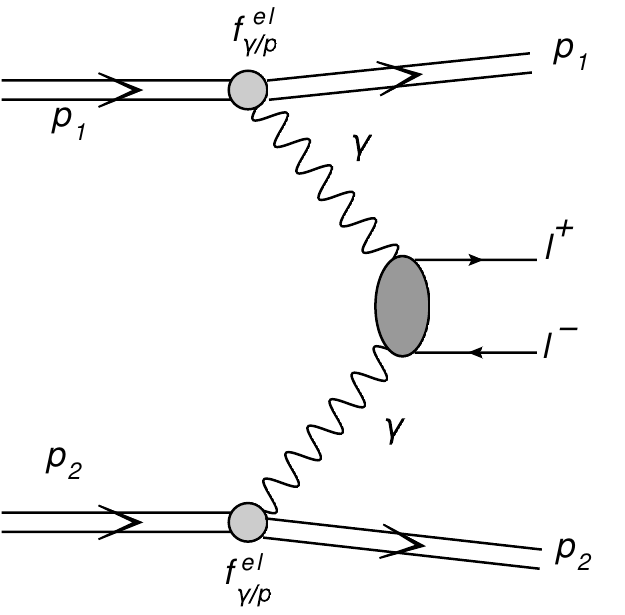}
\includegraphics[width=4cm]{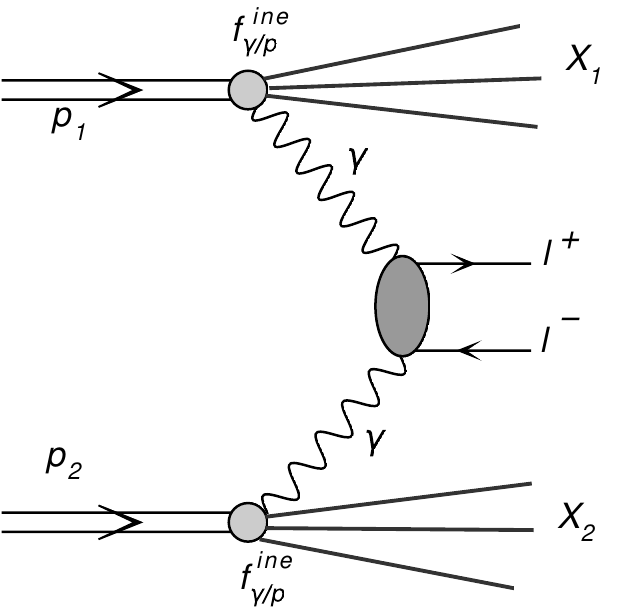}\\
\includegraphics[width=4cm]{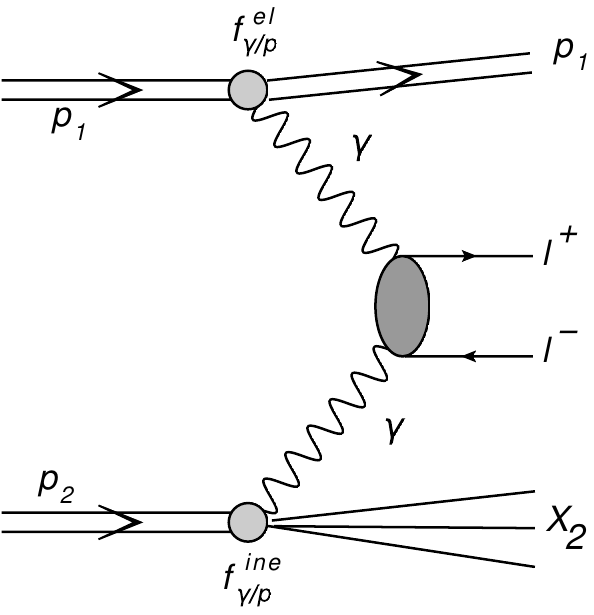}
\includegraphics[width=4cm]{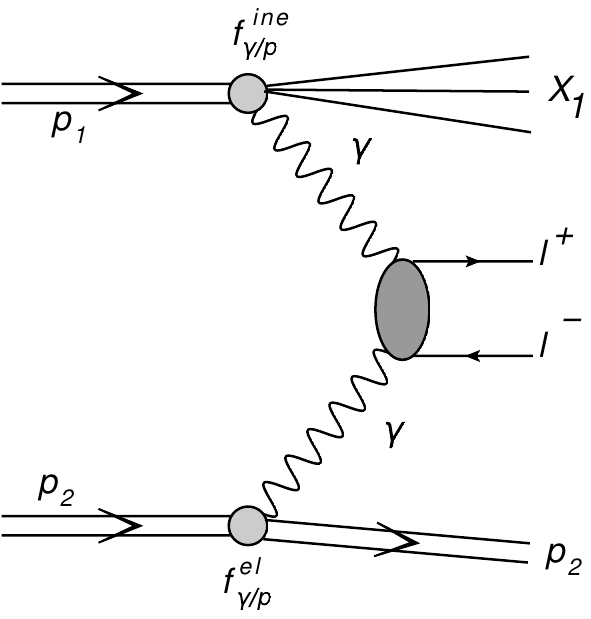}
\caption{Four different categories of $\gamma \gamma$ fusion mechanisms 
of dilepton production in proton-proton collisions.}
\label{fig:diagrams}
\end{center}
\end{figure}

There are four categories of the $\gamma \gamma$ processes
as shown in Fig.\ref{fig:diagrams}. We call them
elastic-elastic, inelastic-inelastic, elastic-inelastic and
inelastic-elastic. The double inelastic contribution is
not included when proton is measured.

In the $k_T$-factorization approach  \cite{SFPSS2015,LSS2016}, 
the cross section for production of $l^+l^-$ can be written in the form
\begin{eqnarray}
{d \sigma^{(i,j)} \over dy_1 dy_2 d^2\bp_1 d^2\bp_2} &&=  
\int  {d^2 \bq_1 \over \pi \bq_1^2} {d^2 \bq_2 \over \pi \bq_2^2}  
{\cal{F}}^{(i)}_{\gamma^*/A}(x_1,\bq_1) \, {\cal{F}}^{(j)}_{\gamma^*/B}(x_2,\bq_2) 
{d \sigma^*(p_1,p_2;\bq_1,\bq_2) \over dy_1 dy_2 d^2\bp_1 d^2\bp_2} \, , \nonumber \\ 
\label{eq:kt-fact}
\end{eqnarray}
where the indices $i,j \in \{\rm{el}, \rm{in} \}$ denote elastic or 
inelastic final states.
Here the photon flux for inelastic case is integrated over the mass
of the remnant.

The ATLAS collaboration analysis imposes
special condition on:
\begin{equation}
\xi_1 = \xi_{ll}^+  \; , \;  \xi_2 = \xi_{ll}^- \; .
\end{equation}
The longitudinal momentum fractions of the photons were calculated
in the ATLAS analysis as:
\begin{eqnarray}
\xi_{ll}^+ &=& \left( M_{ll}/\sqrt{s} \right) \exp(+Y_{ll}) \; , \nonumber \\
\xi_{ll}^- &=& \left( M_{ll}/\sqrt{s} \right) \exp(-Y_{ll}) \; .
\end{eqnarray}
Only lepton variables enter the formula.

\section{Selected results}
\label{sec:results}

\subsection{Our programs}

\begin{figure}
\begin{center}
\includegraphics[width=5cm]{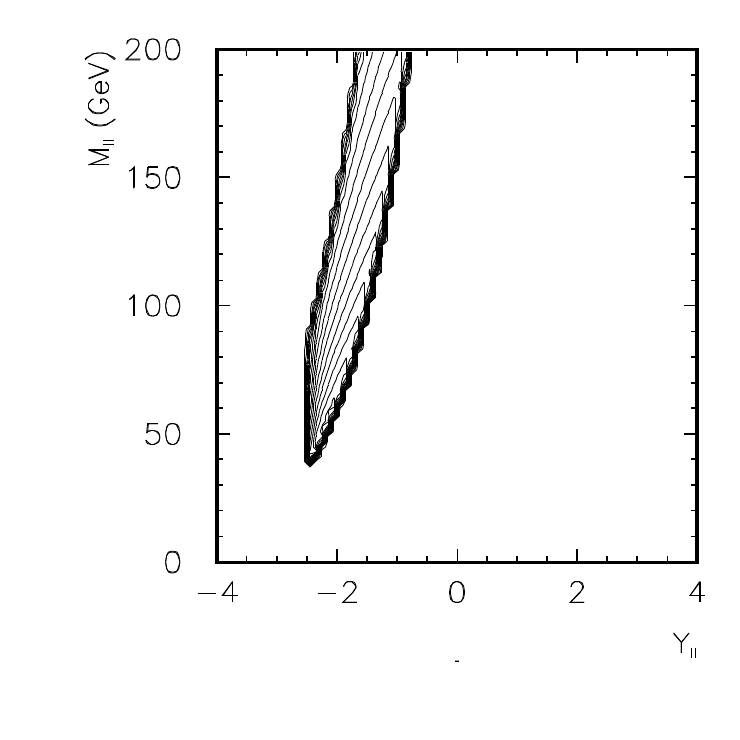}
\includegraphics[width=5cm]{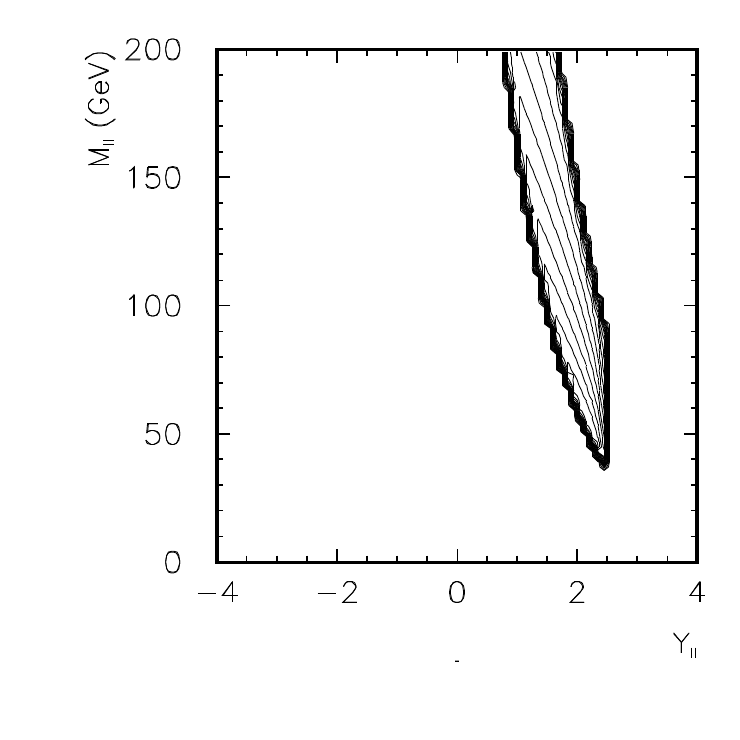}\\ 
\includegraphics[width=5cm]{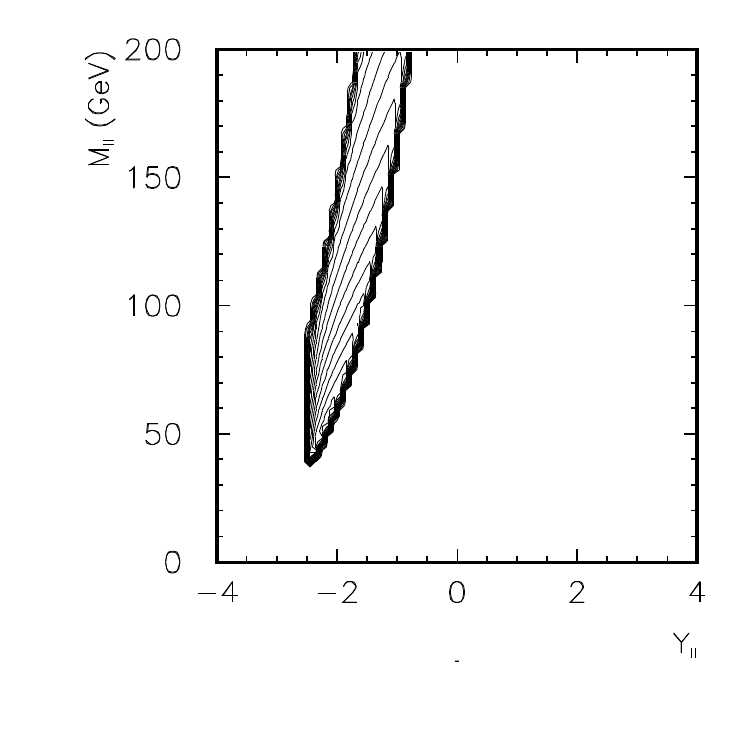}
\includegraphics[width=5cm]{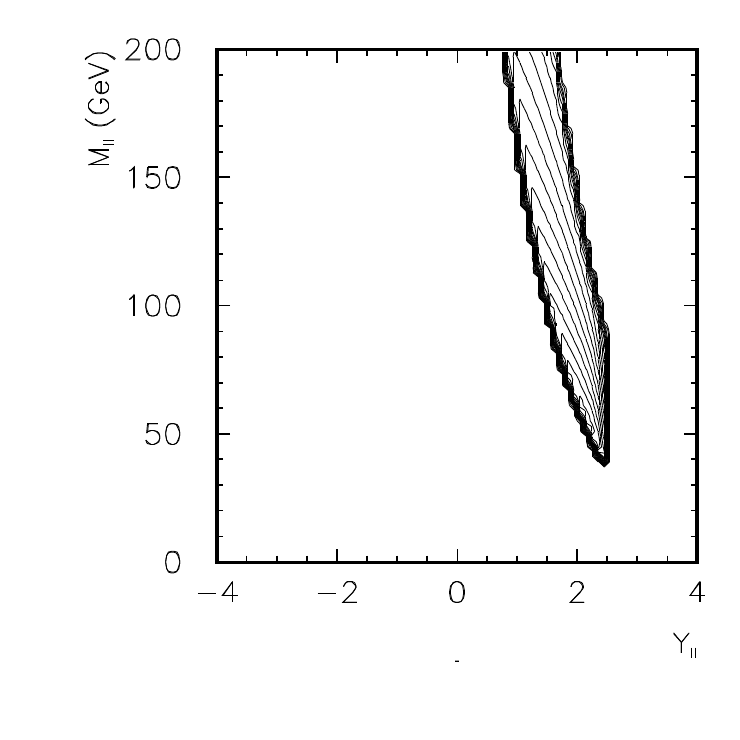}
\caption{
Two-dimensional distribution in ($M_{ll},Y_{ll}$)
for double-elastic contribution (upper rows) and single dissociative
(lower rows).
Here we have imposed experimental condition on $\xi_2$ (left panel) or 
$\xi_1$ (right panel) as explained in the main text.
The $p_{t,\mu} >$ 15 GeV condition was imposed in addition.
The Szczurek-Uleshchenko structure function parametrization was used
here for the single dissociative contribution for illustration.
}
\label{fig:dsig_dMdY_withcuts}
\end{center}
\end{figure}

The measurement of protons has strong influence on many fully leptonic 
observables. 
In Fig. \ref{fig:dsig_dMdY_withcuts} 
we show two-dimensional distributions in ($M_{ll},Y_{ll}$)
for fully elastic (upper panels) and single-dissociative
(lower panels) contributions.
A big part of the phase space is not accessible kinematically which
is related to the cut on $\xi$'s.

\begin{figure}
\begin{center}
\includegraphics[width=5cm]{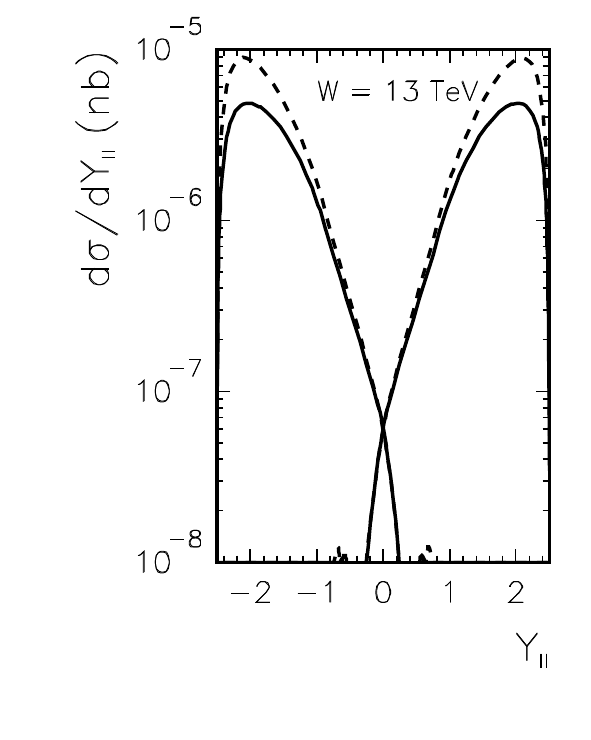}
\caption{Distribution in dilepton rapidity for
four different contributions considered.
Here the cuts on $\xi_{ll}^{+}$ or $\xi_{ll}^-$ are imposed.
The solid line is for double elastic contribution and the dashed line is
for single dissociation contribution.}
\label{fig:dsig_dYll_withcuts}
\end{center}
\end{figure}

In Fig.\ref{fig:dsig_dYll_withcuts} we show a projection on
$Y_{ll}$. One can observe a dip at $Y_{ll} \approx$ 0 which is due to
the imposed cuts. When the cuts are removed the dip is not present
\cite{SLL2021}.

Many other distributions were discussed in \cite{SLL2021}.

\subsection{SuperChic}

In this subsection we show results obtained using the SuperChic-4
generator \cite{HTKR2020}.

\begin{figure}
\begin{center}
\includegraphics[width=6cm]{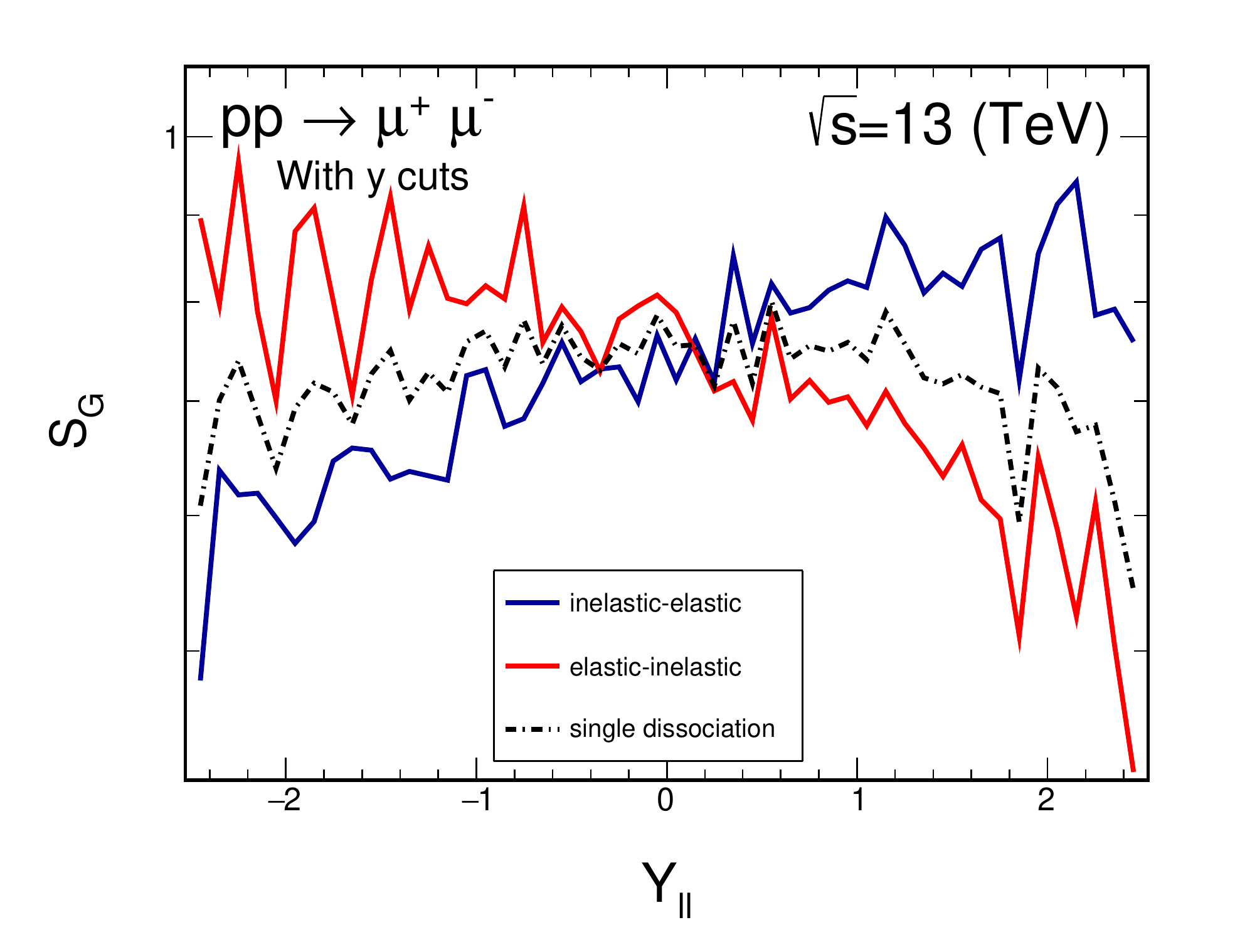}
\includegraphics[width=6cm]{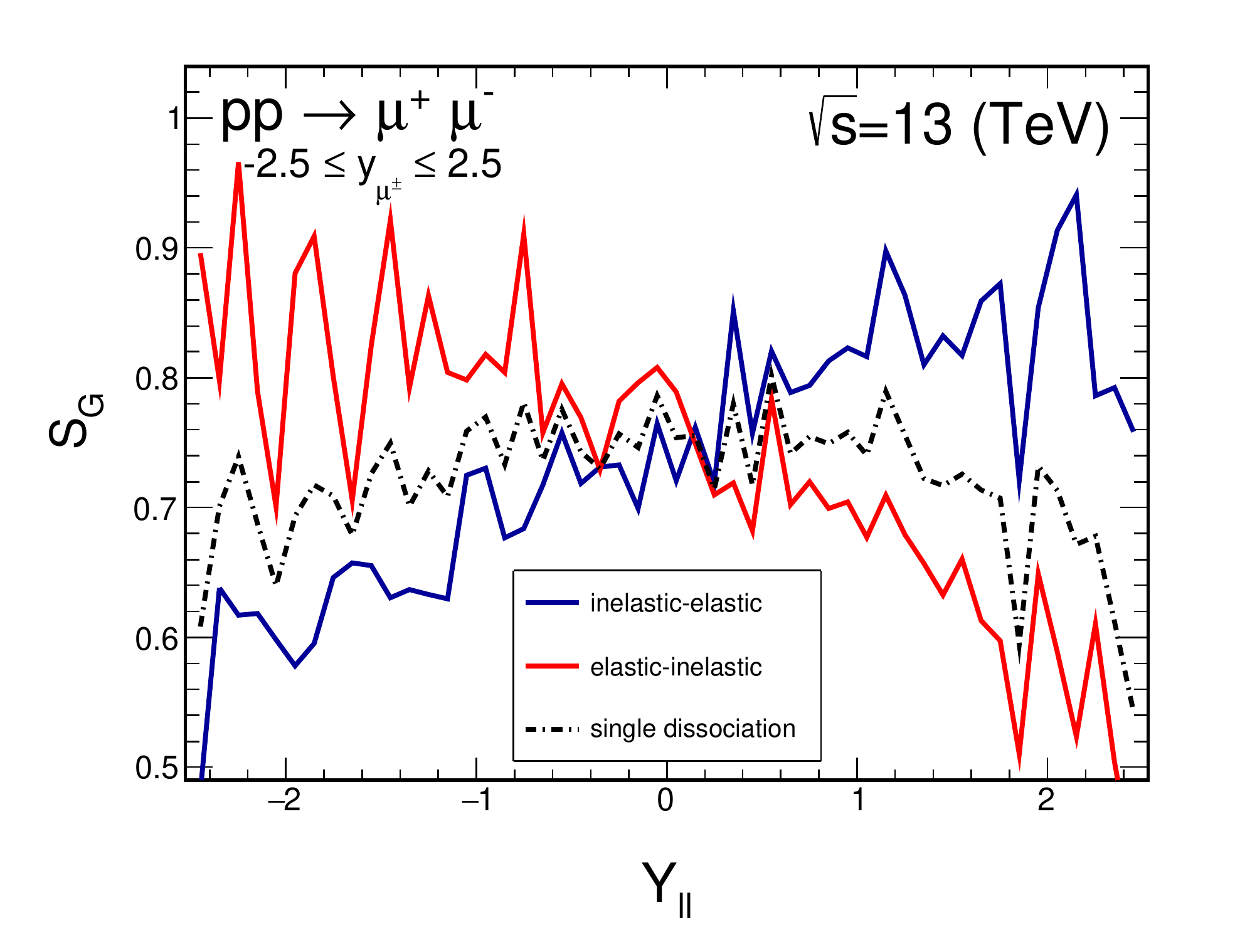}
\caption{The soft gap survival factor as a function of
rapidity of the $\mu^+ \mu^-$ pair for single proton dissociation.
We show the result without $\xi$ cuts (left panel) and
with $\xi$ cuts (right panel). The dash-dotted black line represents
effective gap survival factor for both single-dissociation components
added together.
}
\label{fig:soft_gap_survival_factor_2}
\end{center}
\end{figure}

In Fig.\ref{fig:soft_gap_survival_factor_2} we show corresponding
gap survival factor calculated as:
\begin{eqnarray}
S_G(Y_{ll}) &=& \frac{d \sigma / d Y_{ll}|_{with SR}}
                     {d \sigma / d Y_{ll}|_{without SR}}
\label{differential_gap}
\end{eqnarray}
as a function of $Y_{ll}$ variable.

Without the $\xi$ cut we observe quite different shapes of distributions
in $Y_{ll}$ without and with soft rapidity gap survival factor 
(see the left panel).
When the $\xi$-cut is imposed the distributions with and without
soft rapidity gap survival factor have very similar shapes.
Then, however, the elastic-inelastic and inelastic-elastic
contributions are well separated in $Y_{ll}$.

\begin{figure}
\begin{center}
\includegraphics[width=6cm]{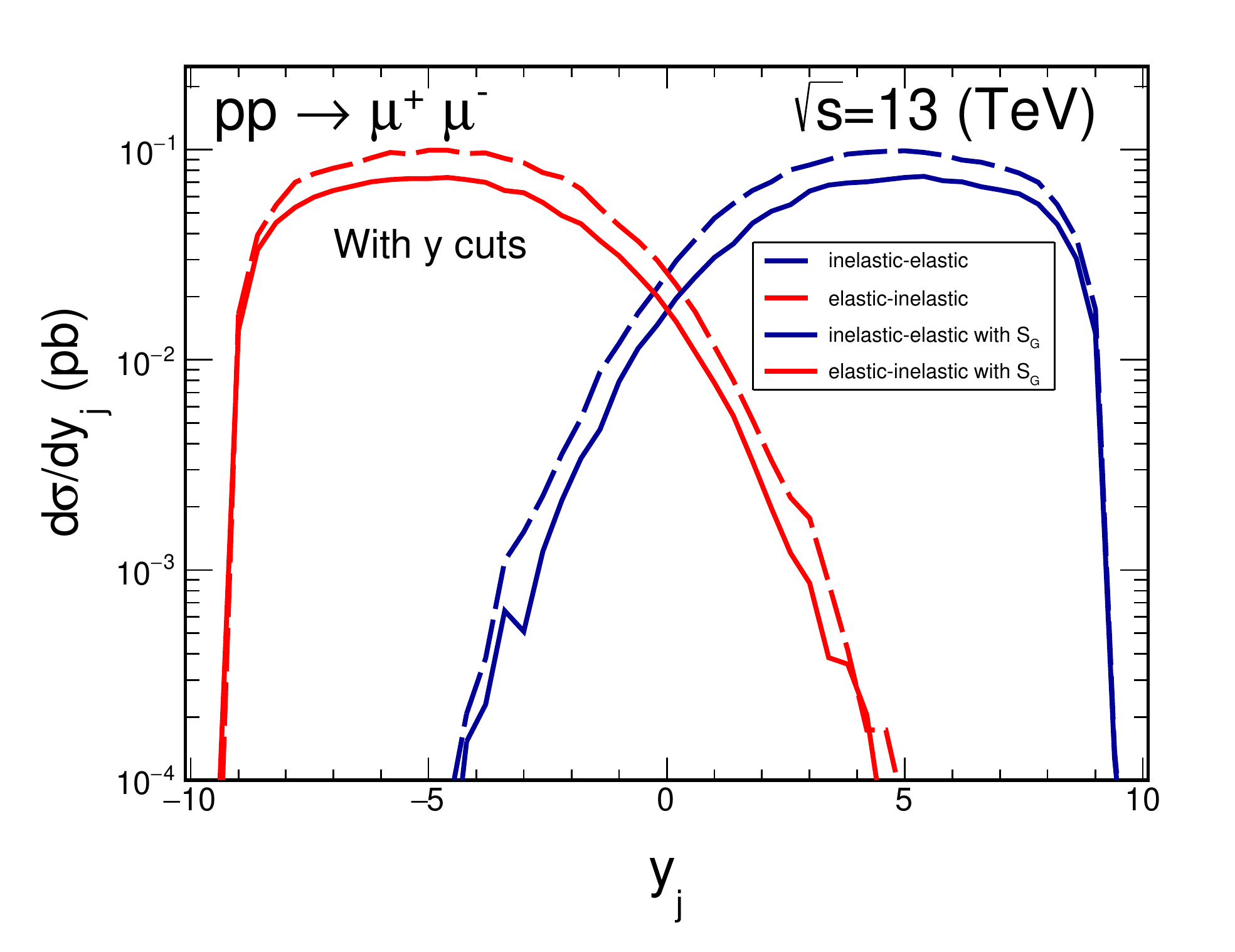}
\includegraphics[width=6cm]{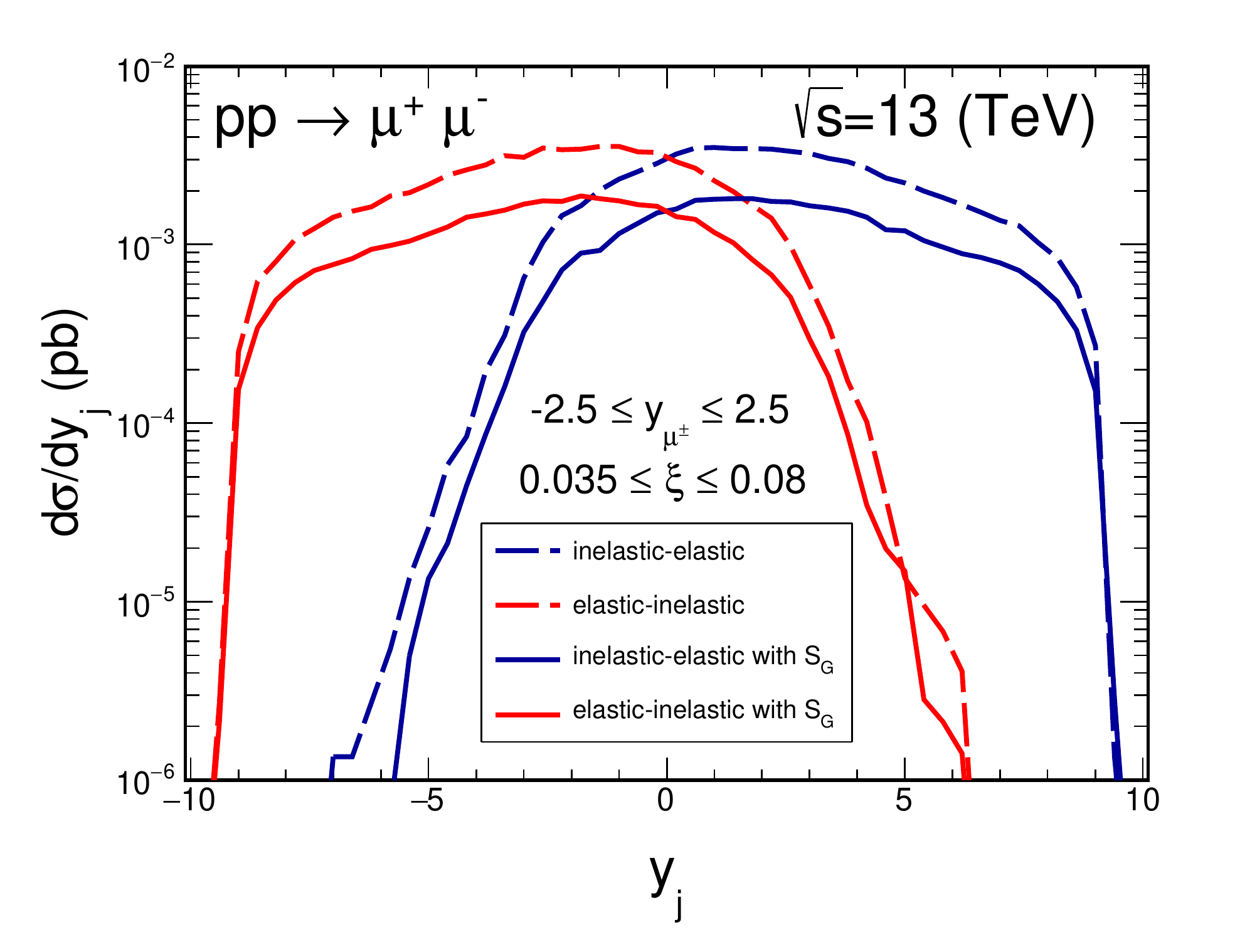}
\caption{Distribution in the (mini)jet rapidity for the inclusive case
with no $\xi$ cut (left panel) and when the cut on $\xi$ is imposed
(right panel) for elastic-inelastic and inelastic-elastic contributions
as obtained from the SuperChic generator.
We show result without (dashed line) and with (solid line) soft
rescattering correction.
}
\label{fig:dsig_dyjet_SUPERCHIC}
\end{center}
\end{figure}

In Fig.\ref{fig:dsig_dyjet_SUPERCHIC} we show the (mini)jet distribution
in rapidity for elastic-inelastic and inelastic-elastic components.
We show the distribution without imposing the $\xi$ cut (left panel)
and when imposing the $\xi$ cut (right panel).
One can observe slightly different shape for both cases.
The corresponding gap survival factor (probability of no jet in the main
detector) is 0.8 and 0.5, respectively.

\section{Conclusion}

Here we have reported our recent studies of $l^+ l^-$ production
in proton-proton scattering with one forward proton, by imposing
a cut on the so-called proton $\xi$ variable.
In this calculation we have included double-elastic and single
dissociative contributions.
In the latter case we have considered both continuum production 
as well as $\Delta^+$ isobar production or production of other 
nucleon resonances, not discussed here explicitly (see \cite{SLL2021}). 

Several distributions were discussed in \cite{SLL2021}.
Here we have shown only some selected results.
Particularly interesting is the distribution 
in $Y_{ll}$ which has a minimum at $Y_{ll} \sim$ 0.
The minimum at $Y_{ll}$ = 0 is caused by the experimental condition on 
$\xi_{ll}^{\pm}$ imposed on the leading proton.

We have also made calculations with the popular SuperChic generator 
and compared corresponding results to the results of our code(s).
In general, the results are very similar to those obtained with
our codes.
We have shown also some results for kinematics-dependent
gap survival factor. We have found some interesting dependence
of gap survival factor on $Y_{ll}$.
Finally we have shown rapidity distribution of a (mini)jet associated
with partonic processes, also when including soft rescattering
corrections.

\section*{Acknowledgements}
This study was partially supported by the Polish National Science Center
grant UMO-2018/31/ /B/ST2/03537 and by the Center for Innovation and
Transfer of Natural Sciences and Engineering Knowledge in Rzesz{\'o}w.



\end{document}